\documentclass[preprint,number]{elsarticle}
\usepackage{amsfonts,amsmath}

\def\kk{{\bf k}}
\begin{document}

\title{A mechanism for pair formation in strongly correlated systems}

\author[ua]{T. Verhulst\corref{cr1}\fnref{fn1}}
\ead{tobias.verhulst@ua.ac.be}

\author[ua]{J. Naudts}
\ead{jan.naudts@ua.ac.be}

\cortext[cr1]{Corresponding author}
\fntext[fn1]{Research Assistant of the Research Foundation - Flanders
(\textsc{fwo} - Vlaanderen)}
\address[ua]{Departement Fysica, Universiteit Antwerpen\\ Groenenborgerlaan 171,
2020 Antwerpen, Belgium}

\date{}

\begin{abstract}
We start from a Hamiltonian describing non-interacting electrons and add bosons
to the model, with a Jaynes-Cummings-like interaction between the bosons and
electrons. Because of the specific form of the interaction the model can be
solved exactly. In the ground state, part of the electrons form bound pairs with
opposite momentum and spin. The model also shows a gap in the kinetic energy of
the electrons, but not in the spectrum of the full Hamiltonian. This gap is not
of a mean-field nature, but is due to the Pauli exclusion principle.
\end{abstract}

\maketitle

\section{Introduction}

In the theory of strongly correlated particles a model known for a long time is
the mean-field model with Hamiltonian
\begin{equation}
 H=-\frac1N\sum_{k\neq
l}\sigma_k^+\sigma_l^-+\epsilon\sum_k\sigma_k^z.\label{eq:meanfield}
\end{equation}
This is the mean-field \textsc{bcs} model \cite{BCS57,BCS57b,BEST61,TW67,TW68}.
Despite the simplicity of this model many interesting results about strongly
coupled particles can be derived from it. Recently, however, there has been a
lot of interest in other models for the creation of paired electrons, motivated
in large part by the problem of finding a theory for high $T_c$
superconductivity \cite{A07,MPS08}.

Among the most useful models for strongly correlated systems are the Hubbard
model and its generalizations \cite{DPS04,BDEE09,H63,EFGKK05}. These models have
been used to describe not only superconductivity \cite{EKS93,D94,JKSKB01} but
also ferromagnetism \cite{MT93}, metal-insulator transitions \cite{M68} and
other properties of solids \cite{KE1994,T98} and of atoms in optical lattices
\cite{D08}.

In this paper we start from a Hamiltonian describing the motion of
non-interacting electrons on a lattice. We then add to the model bosons, which
interact with the electrons in a way similar to the boson-fermion interaction in
the Jaynes-Cummings model \cite{JC63}. The complete Hamiltonian is then similar
to the one used to describe bipolaron interactions \cite{AM94,WRF96,DA09}.
However, in our model this interaction term does not describe the interaction
between pairs of electrons, but rather provides a process which creates and
destroys pairs. A short account of the present findings has already been given
in \cite{NV10}.

In models like (\ref{eq:meanfield}) the binding energy of all pairs is the same.
It has recently been suggested \cite{PC09,PCC09} that, due to the Pauli exclusion
principle, this energy should not be a constant. In our model we find that,
indeed, the binding energy depends on the momenta ${\bf k},-{\bf k}$ of the paired electrons.

Overview of the paper. In section 2 we give a detailed description of the
different terms in the Hamiltonian, both their mathematical properties and
physical interpretations. In section 3 the exact solution of the model for a
fixed value of $\kk,-\kk$ is presented. In section 4 the many particle ground state is
constructed. In section 5 the elementary
excitations and the gap in the energy are discussed, followed by the main
conclusions in section 6. A list of all eigenstates and eigenvalues of the model
can be found in the appendix.

\section{The model}

The Hamiltonian we consider consist of three parts: one describing the electrons,
one describing the bosons and one describing an interaction between them. The
fermion part is given by
\begin{equation}
 H_F=-\hbar\sum_{\bf k,\sigma}\lambda_{\bf k}B^\dagger_{\bf k,\sigma}B_{\bf
k,\sigma}+\eta\hbar\sum_{\kk,\sigma}B^\dagger_{\kk,\sigma}B_{\kk,\sigma}B^\dagger_{
-\kk\sigma}B_{-\kk,\sigma}+\frac{\hbar}{2}\sum_{\bf k}\omega_{\bf k}J^z_\kk.
\label{eq:Hfermi}
\end{equation}
The first term comes from the kinetic energy
$T=-\sum_{i,j}t_{ij}\sum_{\sigma}b^\dagger_{i,\sigma}b_{j,\sigma}$ which is
written in diagonal form using Bogoliubov's quasi-particles. The quasi-particle
operators are defined by $B_{\bf k,\sigma}=\sum_iv_{{\bf k},i}b_{i,\sigma}$ with
$v_{{\bf k},i}$ the matrix coefficients of the diagonalizing unitary
transformation. These operators satisfy the canonical anti-commutation
relations. The $\lambda_{\bf k}$ are the eigenvalues of the kinetic energy. On a
cubic $N\times N\times N$ lattice they are given by
\begin{equation}
 \lambda_{\bf k}=2\sum_{\alpha=1}^3\cos(2\pi k_\alpha)\label{eq:cubic}
\end{equation}
where $k_\alpha=\frac{2n-N+1}{2N}$ for $n$ from $0$ to $N-1$.

The second part of $H_F$ describes an interaction between electrons with equal
spin. If the parameter $\eta$ is positive, this term raises the energy of
states where electrons with opposite momenta have equal spins.

The operator $J^z_{\bf k}$ in the last term of (\ref{eq:Hfermi}) is defined as
$J^z_{\bf k}=\frac12[J^+_{\bf k},J^-_{\bf k}]$ with $J^-_{\bf k}=B^\dagger_{-\bf
k,\uparrow}B^\dagger_{\bf k,\downarrow}B_{-\bf k,\downarrow}B_{\bf k,\uparrow}$
and $J^+_{\bf k}=(J^-_{\bf k})^\dagger$. $J^z_{\bf k}$ and $J^\pm_{\bf k}$
together generate a representation of $\mathfrak{su}(2)$ and can be interpreted
as some kind of spin operators. An expression for $J^z_{\bf k}$ in counting
operators $n_{\bf k,\sigma}=B^\dagger_{\bf k,\sigma}B_{\bf k,\sigma}$ can be
found:
\begin{equation}
 J^z_{\bf k}=\frac12n_{\bf k,\uparrow}(1-n_{-\bf k,\uparrow})n_{-\bf
k,\downarrow}(1-n_{\bf k,\downarrow})-\frac12n_{\bf k,\downarrow}(1-n_{-\bf
k,\downarrow})n_{-\bf k,\uparrow}(1-n_{\bf k,\uparrow}).
\end{equation}
It is then easy to see that $J^z_{\bf k}$ has eigenvalues $-\frac12$, $0$, and
$\frac12$, as one would expect. This part of the Hamiltonian describes the
interaction with a constant external field.

Notice that one could add some further terms quadratic in the $B_{\bf k,\sigma}$'s.
For instance, we do not consider a spin flip part in the Hamiltonian, or terms changing the total
momentum. We could add a term proportional to $\sum_{\bf k,\sigma}B^\dagger_{\bf
k,\sigma}B_{-\bf k,\sigma}$. This would make the calculations more
complicated, although the model remains integrable. However, such a term does
not change anything fundamentally. So we neglect it here.

The bosonic part of the Hamiltonian describes a collection of harmonic oscillators,
one for each fermionic degree of freedom. The creation and
annihilation operators are denoted $c_{{\bf k},\tau}$ and $c^\dagger_{{\bf k},\tau}$,
with $\tau=\uparrow$ or $=\downarrow$. For simplicity of notation we use the
same labels as for the fermionic operators. These oscillators could for instance be
lattice vibrations with two possible polarizations.
The bosonic Hamiltonian reads
\begin{equation}
 H_B=\sum_{\bf k}\hbar\mu_{\bf k}\left(c^\dagger_{{\bf k},\uparrow}c_{{\bf
k},\uparrow}+c^\dagger_{{\bf k},\downarrow}c_{{\bf k},\downarrow}\right)
=\sum_{\bf k}\hbar\mu_{\bf k}\sum_{\tau=\uparrow,\downarrow}m_{{\bf k},\tau},
\end{equation}
with $\mu_{\bf k}=\mu_{-\bf k}>0$ and with
 $m_{{\bf k},\tau}=c^\dagger_{{\bf k},\tau}c_{{\bf k},\tau}$.

By introducing new operators $a_{\bf k}=c_{{\bf k},\uparrow}c^\dagger_{-{\bf k},\downarrow}$
this bosonic Hamiltonian can be rewritten as
\begin{equation}
 H_B=\sum_{\bf k}\hbar \mu_{\bf k}\left(a^\dagger_{\bf k}a_{\bf k}
+a_{\bf k}a^\dagger_{\bf k}\right)-\sum_{\bf k}\hbar\mu_{\bf k}m_{\bf k,\uparrow}m_{-\bf k,\downarrow}.
\label{eq:Hbose}
\end{equation}
While the operators $c_{\bf{k},\tau}$ and $c^\dagger_{\bf{k},\tau}$ satisfy the
canonical commutation relations for bosons, the $a_{\bf k}$ and $a^\dagger_{\bf
k}$ operators do not. It still holds that $[a_{\bf k},a_{\bf l}]=0$, but the
commutation between the new creation and annihilation operators is more
complicated:
\begin{equation}
 [a_{\bf k},a^\dagger_{\bf l}]=\delta_{{\bf k},{\bf l}}\left(-c_{{\bf
k},\uparrow}c^\dagger_{{\bf k},\uparrow}+c_{-{\bf k},\downarrow}c^\dagger_{-{\bf
k},\downarrow}\right)
=-\delta_{{\bf k},{\bf l}}\left(m_{\bf k,\uparrow}-m_{-\bf k,\downarrow}
\right).
\end{equation}
The operators $a_{\bf k}$ and $a^\dagger_{\bf k}$ are introduced because they
are useful to express the interaction term (\ref{eq:Hint}), which will now be introduced.

The interaction between the electrons and bosons is described by the Hamiltonian
\begin{equation}
 H_I=\hbar\sum_{\bf k}\xi_{\bf k}\left(a^\dagger_{\bf k}J^-_{\bf k}+a_{\bf
k}J^+_{\bf k}\right)\label{eq:Hint}
\end{equation}
with $a_{\bf k}$, $a^\dagger_{\bf k}$ and $J^\pm_{\bf k}$ as defined above. This
interaction is of fourth order in the fermionic operators $B_{\bf k,\sigma}$ and
$B_{\bf k,\sigma}^\dagger$, and of second order in the bosonic operators $c_{\bf
k,\sigma}$ and $c_{\bf k,\sigma}^\dagger$.

The complete Hamiltonian $H=H_F+H_B+H_I$ describes electrons moving on a lattice and
interacting with each other both directly and mediated by lattice vibrations.
Notice that the latter interaction is of the Jaynes-Cummings type and is
expressed in the operators $J_{\bf k}$ and $a_{\bf k}$, not in the original creation
and annihilation operators
$B_{{\bf k},\sigma}$ and $c_{{\bf k},\sigma}$.

We can compare the present model with the mean-field model
(\ref{eq:meanfield}) by identifying $J_\kk^-$ with $\sigma_\kk^+\sigma_{-\kk}^-$. Here,
there are only interactions between particles when they
have the same or opposite momentum $\pm \bf k$,
while in (\ref{eq:meanfield}) there is a mean-field interaction between
particles with all possible momenta. In addition, the interaction term of our model is of
second order in the $\sigma_\kk$
\begin{equation}
H_I=\hbar\sum_\kk\xi_\kk\left(a^\dagger_{\kk}\sigma_\kk^+\sigma_{-\kk}^-+a_{\kk}\sigma_{-\kk}
^+\sigma_\kk^-\right).
\end{equation}

Thus, integrating out the bosons gives a term of fourth
order in the $\sigma_\kk$ instead of one of second order.

\section{The exact solution of the model}

Because there is no interaction between particles with momenta $\bf k$
respectively ${\bf k}'$ unless ${\bf k}=\pm{\bf k}'$, we construct solutions in
a subspace $\mathcal H_\kk$ with only particles with the same or opposite momenta.
Thus, the number of electrons is between zero and four while the number of
bosons is between zero and infinity.

We use the following basis vectors:
\begin{equation}
 \left|n_{\kk,\uparrow},m_{\kk,\uparrow};n_{\kk,\downarrow},m_{\kk,\downarrow};n_{-\kk,
\uparrow},m_{-\kk,\uparrow};n_{-\kk,\downarrow},m_{-\kk,\downarrow}\right\rangle\label
{eq:basis}
\end{equation}
where $n=0$ or $1$ is the number of electrons for the given momentum and spin, and $m$
counts the number of corresponding bosons. There are two important things to
note about these basis vectors. First: most terms of the Hamiltonian are already
diagonal in this basis. Only the fermion-boson interaction $H_I$ is not. Second:
on the vectors where it is not zero, the interaction Hamiltonian $H_I$ acts like
a Jaynes-Cummings Hamiltonian. This means that the well known techniques to
solve the Jaynes-Cummings model can be applied.

\subsection{The diagonal parts of $H$}

Most terms of $H$ are diagonal in the basis (\ref{eq:basis}). The eigenvalues
are given here. For the kinetic energy of the electrons
\begin{equation}
 T=-\sum_{i,j}t_{ij}\sum_{\sigma}b^\dagger_{i,\sigma}b_{j,\sigma}=-\hbar\sum_{
\bf k,\sigma}\lambda_{\bf k}B^\dagger_{\bf k,\sigma}B_{\bf k,\sigma}
\end{equation}
the eigenvalues are
\begin{equation}
 -\hbar\lambda_{\kk}(n_{\kk,\uparrow}+n_{\kk,\downarrow})-\hbar\lambda_{-\kk}(n_{
-\kk\uparrow}+n_{-\kk,\downarrow}).
\end{equation}
The term $\frac{\hbar}{2}\sum_{\bf k}\omega_{\bf k}J^z_{\bf k}$ has eigenvalues
\begin{multline}
\frac{\hbar(\omega_\kk+\omega_{-\kk})}{2}\left[n_{\kk,\uparrow}(1-n_{-\kk,\uparrow})n_{
-\kk,\downarrow}(1-n_{\kk,\downarrow})\right.\\\left.-n_{\kk,\downarrow}(1-n_{-\kk,
\downarrow})n_{-\kk,\uparrow}(1-n_{\kk,\uparrow})\right].
\end{multline}
The second term of $H_F$, the fermion-fermion interaction, has eigenvalues
\begin{equation}
 \eta\hbar\left(n_{\kk,\uparrow}n_{-\kk,\uparrow}+n_{\kk,\downarrow}n_{-\kk,\downarrow}
\right).
\end{equation}
The eigenvalues of the bosonic Hamiltonian $H_B$ are
\begin{equation}
\mu_\kk\left(m_{\kk,\uparrow}+m_{-\kk,\downarrow}\right).
\end{equation}
All eigenstates and eigenvalues are listed in the appendix.

\subsection{The interaction Hamiltonian $H_I$}

The interaction Hamiltonian $H_I$ vanishes when the number of electrons is not equal to two.
In addition, straightforward calculation shows that it vanishes when the following conditions are satisfied
\begin{enumerate}
 \item $n_{\kk,\uparrow}=0$ or $n_{-\kk,\uparrow}=0$ or $m_{-\kk,\downarrow}=0$,
 \item $n_{-\kk,\uparrow}=0$ or $n_{\kk,\uparrow}=0$ or $m_{\kk,\uparrow}=0$,
 \item $n_{-\kk,\uparrow}=0$ or $n_{\kk,\uparrow}=0$ or $m_{\kk,\downarrow}=0$,
 \item $n_{\kk,\uparrow}=0$ or $n_{-\kk,\uparrow}=0$ or $m_{-\kk,\uparrow}=0$.
\end{enumerate}
In particular, the only basis vectors on which $H_I$ does not vanish are the two-fermion vectors with vanishing momentum and
vanishing spin
\begin{eqnarray}
 &&\left|1,m_{\kk,\uparrow};0,m_{\kk,\downarrow};0,m_{-\kk,
\uparrow};1,m_{-\kk,\downarrow}\right\rangle\\
 &&\left|0,m_{\kk,\uparrow};1,m_{\kk,\downarrow};1,m_{-\kk,
\uparrow};0,m_{-\kk,\downarrow}\right\rangle.
\end{eqnarray}
The eigenstates of $H$ can now be constructed in a way similar to the solution
of the Jaynes-Cummings model. Although it is not needed in order to obtain the
eigenvalues and eigenvectors we assume from now on that $\lambda_\kk=\lambda_{-\kk}$,
as is for instance the case for a cubic lattice. This assumption
simplifies the calculations a lot.
The eigenvectors on which $H_I$ does not vanish are denoted as follows
\begin{eqnarray}
 |2,0,+\kk,+\rangle&=&\sin\theta_\kk^+|1,m_{\kk,\uparrow};0,m_{\kk,\downarrow};0,m_
{-\kk,\uparrow};1,m_{-\kk,\downarrow}+1\rangle\nonumber\\
 &&\cos\theta_\kk^+|0,m_{\kk,\uparrow}+1;1,m_{\kk,\downarrow};1,m_{-\kk,
\uparrow};0,m_{-\kk,\downarrow}\rangle
 \\
 |2,0,+\kk,-\rangle&=&\cos\theta_\kk^+|1,m_{\kk,\uparrow}+1;0,m_{\kk,\downarrow};0,
m_{-\kk,\uparrow};1,m_{-\kk,\downarrow}\rangle\nonumber\\
 &&-\sin\theta_\kk^+|0,m_{\kk,\uparrow};1,m_{\kk,\downarrow};1,m_{-\kk,
\uparrow};0,m_{-\kk,\downarrow}+1\rangle
 \\
 |2,0,-\kk,+\rangle&=&\sin\theta_\kk^-|0,m_{\kk,\uparrow};1,m_{\kk,\downarrow}
;1,m_{-\kk,\uparrow};0,m_{-\kk,\downarrow}+1\rangle\nonumber\\
 &&\cos\theta_\kk^-|1,m_{\kk,\uparrow}+1;0,m_{\kk,\downarrow};0,m_{-\kk,\uparrow}
;1,m_{-\kk,\downarrow}\rangle
 \\
 |2,0,-\kk,-\rangle&=&\cos\theta_\kk^-|0,m_{\kk,\uparrow}+1;1,m_{\kk,\downarrow}
;1,m_{-\kk,\uparrow};0,m_{-\kk,\downarrow}\rangle\nonumber\\
 &&-\sin\theta_\kk^-|1,m_{\kk,\uparrow};0,m_{\kk,\downarrow};0,m_{-\kk,\uparrow}
;1,m_{-\kk,\downarrow}+1\rangle.
\end{eqnarray}
In these expressions the angles $\theta_\kk^\pm$ are given by
\begin{equation}
 \theta_\kk^\pm=\arctan\left(\frac{\omega_\kk+\omega_{-\kk}}{2\xi_\kk}\mp\frac{1}{
2\xi_\kk}\sqrt{(\omega_\kk+\omega_{-\kk})^2+4\xi_\kk^2}\right).
\end{equation}
The eigenvalues are $\varepsilon_2^+$ for the first and the third class,
and $\varepsilon_2^-$ for the second and the fourth class, with
\begin{equation}
 \varepsilon_2^{\pm}=E_{HO}^\kk-2\hbar\lambda_\kk\pm\frac{\hbar}{2}\sqrt{
(\omega_\kk+\omega_{-\kk})^2+4\xi_\kk^2}.
\end{equation}
The energy $E_{HO}^\kk$ of the harmonic oscillators equals
\begin{equation}
 E_{HO}^\kk=m_{\kk,\uparrow}+m_{\kk,\downarrow}+m_{-\kk,\uparrow}+m_{-\kk,\downarrow}.
\label {EHO}
\end{equation}
Note that at least one of the harmonic oscillators is in an excited state.
Thus, the contribution from the harmonic oscillators does not vanish.

\subsection{The ground state for fixed $\kk$}

First we construct the ground state for a fixed value of $\kk$. Since particles
do not interact except when they have equal or opposite momenta,
the complete many-particle ground state
can be constructed from the ground states for fixed $\kk$. We assume from now on that there is
no external field, so $\omega_{\pm \kk}=0$ in (\ref {eq:Hfermi}). The eigenvalues for the
Jaynes-Cummings-like states then become
\begin{equation}
 \epsilon_2^{\pm}=E_{HO}^\kk-2\hbar\lambda_\kk\pm\hbar\xi_\kk.
\end{equation}
For simplicity let us assume now that $\xi_\kk>0$.
If it is large enough then one of the
$\epsilon_2^{-}$ is the lowest energy level. This requires that the
contribution from the bosons is minimal, this is, equal to $E_{HO}^\kk=\hbar\mu_\kk$. Even
then the ground state is still
degenerate because of the symmetry between the momenta $\kk$ and $-\kk$. Indeed, for
large enough $\xi_\kk$, the ground states are
\begin{eqnarray}
  |2,0,\kk,+\rangle&=&\cos\theta_\kk^+|1,1;0,0;0,0;1,0\rangle
 -\sin\theta_\kk^+|0,0;1,0;1,0;0,1\rangle\\
  |2,0,-\kk,+\rangle&=&\cos\theta_\kk^-|0,1;1,0;1,0;0,0\rangle
 -\sin\theta_\kk^-|1,0;0,0;0,0;1,1\rangle.
\end{eqnarray}
If $\xi_\kk$ is small the ground state is the unique state with four electrons and
no bosons
\begin{equation}
 |4,0,0\rangle=|1,0;1,0;1,0;1,0\rangle
\end{equation}
This eigenstate has eigenvalue $\epsilon_4=-4\hbar\lambda_\kk+2\hbar\eta$. The critical
value of $\xi_\kk$ at which the quantum phase transition occurs is
\begin{equation}
 \xi_\kk=\mu_\kk+2(\lambda_\kk-\eta_\kk).\label{eq:critic}
\end{equation}
The second term comes from the difference between $\epsilon_2^-$ and
$\epsilon_4$ with the same number of bosons while the first term stems from the
fact that for the interacting states $E_{HO}^\kk$ can not vanish.

\subsection{The many-particle ground state}

Substituting (\ref{eq:cubic}) in (\ref{eq:critic}) gives:
\begin{equation}
 \xi_\kk=\mu_\kk+2\left(2\sum_{\alpha=1}^3\cos(2\pi k_\alpha)-\eta\right)
\end{equation}
or
\begin{equation}
 \sum_{\alpha=1}^3\cos(2\pi
k_\alpha)=\frac{\xi_\kk}{4}-\frac{\mu_\kk}{4}+\frac{\eta}{2}.
\end{equation}
From this equation one can, at least in principle, determine at which $\kk$ it
becomes energetically favorable to have a ``pair" of electrons bound by a boson
instead of four non-interacting electrons.

In order to do this one needs, in addition to the assumption (\ref{eq:cubic})
for $\lambda_\kk$, expressions for $\xi_\kk$ and $\mu_\kk$. For the interaction
strength $\xi_\kk$ it is reasonable to assume no dependence on $\kk$. We do not
specify an explicit dispersion relation for the bosons.

\section{Filling the band}

The number of electrons which occupy a certain $(\kk,-\kk)$-level varies between zero and
four. If the interaction strength is zero, all electrons
occupy levels with four electrons and zero bosons. However, with a nonzero
interaction part of the electrons form pairs. Thus, for
some $(\kk,-\kk)$-levels there will be two electrons and one boson instead of four
electrons and no bosons. This means that, if the number of electrons stays the
same, some of them must occupy higher $\kk$-levels. This effect inhibits to some
extend the formation of pairs because some of the pairs have a higher kinetic energy.

We describe in detail the case of a one dimensional system with a band filled up
to the Fermi level at $k=k_F$. We denote $\lambda_{k_F}\equiv\lambda_F$ and
$\mu_{k_F}\equiv\mu_F$. $|4,0,0\rangle_{k_F}$ is then the highest occupied level
with energy $4\hbar(\eta-\lambda_F)$. If these electrons occupy two levels with two electrons each
instead of one level with four electrons, then one pair has $k=k_F$ (the level
$|2,0,k,-\rangle$) and the other one $k=k_F+\delta k$ (the level $|2,0,k+\delta
k,-\rangle$). The pair at $k=k_F$ has the energy
\begin{equation}
 \epsilon_2^-=\hbar\mu_F+2\hbar(\eta-\lambda_F)-\hbar\xi
\end{equation}
and the pair at $k=k_F+\delta k$ has the energy
\begin{equation}
 \hbar\mu_{k_F+\delta k}+2\hbar(\eta-\lambda_{k_F+\delta k})-\hbar\xi.
\end{equation}
Thus, the energy gained by creating two pairs from four free electrons equals
\begin{equation}
 \Delta\epsilon(k_F)=2\hbar(\lambda_{k_F+\delta
k}-\lambda_F+\xi)-\hbar(\mu_F+\mu_{k_F+\delta k})
\label{eq:DeltaE}
\end{equation}
Around the Fermi level one can make expansions for $\lambda_k$ and $\mu_k$
\begin{eqnarray}
 \lambda_k=\lambda_F+\frac{d\lambda_k}{dk}(\lambda_F)\delta k+\left((\delta
k)^2\right)\\
 \mu_k=\mu_F+\frac{d\mu_k}{dk}(\mu_F)\delta k+\left((\delta k)^2\right).
\end{eqnarray}
Substituting these expansions in (\ref{eq:DeltaE}) gives
\begin{equation}
 \Delta\epsilon(k_F)=\hbar(2\xi-\mu_F)+\hbar\left(2\frac{d\lambda_k}{dk}
(\lambda_F)-\frac{d\mu_k}{dk}(\mu_F)\right)\delta k.
\end{equation}
Now the four electrons occupying a second $k$-level can be split into two pairs.
The highest four fermion level is the one with $k=k_F-\delta k$. The
pairs are formed at $k=k_F-\delta k$ and $k=k_F+2\delta k$. The gain in energy
is
\begin{equation}
 \Delta\epsilon(k_F-\delta
k)=\hbar(2\xi-\mu_F)+\hbar\left(6\frac{d\lambda_k}{dk}(\lambda_F)-\frac{
d\mu_k}{dk}(\mu_F)\right)\delta k.
\end{equation}
The same procedure where the particles from a four fermion level at
$k=k_F-n\delta k$ are combined in two paired states at $k=k_F-n\delta k$ and
$k=k_F+(n+1)\delta k$ can now be repeated $N$ times. $N$ can be determined from
the condition
\begin{equation}
 \Delta\epsilon(k_F-N\delta
k)=\hbar(2\xi-\mu_F)+\hbar\left(2(2N+1)\frac{d\lambda_k}{dk}
(\lambda_F)-\frac{d\mu_k}{dk}(\mu_F)\right)\delta k=0.
\end{equation}
Some straightforward algebra then shows that the number of pairs formed is
\begin{equation}
 2N+1=\frac{\frac{d\mu_k}{dk}(\mu_F)}{2\frac{d\lambda_k}{dk}(\lambda_F)}+\frac{
\mu_F-2\xi}{2\delta k\frac{d\lambda_k}{dk}(\lambda_F)}.\label{eq:number}
\end{equation}
Notice that, all other things being equal, a smaller $\delta k$ implies a larger
$N$. Thus, the closer the possible values of $k$ are to each other the more
pairs are formed. In general, $\delta k$ is inversely proportional to the size of
the system, so the fraction of pair-forming electrons is constant as a function
of the system size.

\section{Excitations and kinetic energy gap}

The excitations with the lowest energy are the production of two fermion pairs
at $k=k_F-(N+1)\delta k$ and $k=k_F+(N+2)\delta k$ and the recombination of two
pairs into one four fermion level at $k=k_F-N\delta k$. During the creation of
two pairs from the particles at $k=k_F-(N+1)\delta k$, the electrons in the pair
with the highest $k$-value each get an additional momentum of $(2N+3)\delta k$.
The momentum of the pair with the lowest $k$-value does not change. Thus, the
total change of momentum is
\begin{equation}
 \Delta k_{tot}=2(2N+3)\delta k.
\end{equation}
One concludes that there is a gap in the kinetic energy of the electrons. Substituting
(\ref{eq:number})  gives
\begin{equation}
 \Delta
k_{tot}=\left(\frac{\frac{d\mu_k}{dk}(\mu_F)}{\frac{d\lambda_k}{dk}(\lambda_F)}
+1\right)\delta
k+\frac{\mu_F-\xi}{\frac{d\lambda_k}{dk}(\lambda_F)}.\label{eq:moment}
\end{equation}
Notice that the second term does not depend on $\delta k$, so even if $\delta k$
is very small $\Delta k_{tot}$ may still be macroscopic. An analogous reasoning
holds for the reversed process in which two pairs are destroyed.

\section{Conclusions}

In summary, we start from a free hopping Hamiltonian for the electrons to
develop a model with both electrons and bosons. The interaction between the bosons
and the electrons is inspired by the interaction term in the
Jaynes-Cummings model.

Due to the interaction term part of the electrons become bound in
pairs with opposite momentum and spin. In general, the many particle ground
state contains both paired and unpaired electrons. Equation (\ref{eq:number})
shows in the one-dimensional case
how the number of pairs formed in the ground state depends on the
parameters of the model.

We also demonstrate the existence of a gap in the kinetic energy spectrum.
The gap in this model is not of a mean-field type but is a
consequence of the Pauli exclusion principle. We also find that the binding
energy of the pairs is not a constant but depends on the wavevector $\kk$. This is in agreement
with earlier results\cite{PC09,PCC09} which suggest that, due to the Pauli
exclusion principle, the binding energy of Cooper pairs depends on $\kk$.
In the one-dimensional case, the
total change in momentum for the elementary excitation in which two additional
pairs are formed is given by (\ref{eq:moment}).

Future research is needed to study the thermodynamical properties of the
system. Note that it is possible to write our model as a
mean-field model similar to (\ref{eq:meanfield}). This can be done by first
integrating out the bosons, which gives a Hamiltonian of fourth order in the
$B_\kk$ operators, and then introducing new operators $d^\dagger_\kk$ and $d_\kk$
which create and annihilate two bound pairs. Using these new operators an
effective Hamiltonian can be constructed, similar to (\ref{eq:meanfield}).
However, this new mean-field model still allows the binding energy of the pairs
to depend on $\kk$, which is not the case in (\ref{eq:meanfield}).

\appendix

\section{The eigenstates and eigenvalues}

Here we give an overview of all eigenstates and eigenvalues of the model. The
notation is as follows: the first number counts the electrons and the second
number is the total spin. 
The third label is the total momentum of the electrons. The quantum numbers of the
bosons are omitted.
In one case additional quantum numbers are needed.

We further use the notation $E_{HO}^\kk$ for the eigenvalues of the boson Hamiltonian for one $\kk$,
see (\ref {EHO}).
In particular, the eigenvalues of $H_B$ have the form $\sum_\kk E_{HO}^\kk$.

\begin{enumerate}
\item The eigenvectors with vanishing electron count are of the form
\begin{eqnarray}
|0,0,0\rangle&\equiv&
 |0,m_{\kk,\uparrow};0,m_{\kk,\downarrow};0,m_{-\kk,\uparrow};0,m_{-\kk,\downarrow}\rangle.
\end{eqnarray}
The energy is $\varepsilon_0=E_{HO}^\kk$, this is, only the harmonic oscillators contribute.

\item With one electron there are four classes of states because the electron can have either
momentum $\kk$ or $-\kk$. They are
\begin{eqnarray}
 |1,+\frac12,+\kk\rangle&\equiv&|1,m_{\kk,\uparrow};0,m_{\kk,\downarrow};0,m_{-\kk,\uparrow}
;0,m_{-\kk,\downarrow}\rangle\nonumber,\\
 |1,-\frac12,+\kk\rangle&\equiv&|0,m_{\kk,\uparrow};0,m_{\kk,\downarrow};0,m_{-\kk,\uparrow}
;1,m_{-\kk,\downarrow}\rangle,\\
 |1,+\frac12,-\kk\rangle&\equiv&|0,m_{\kk,\uparrow};1,m_{\kk,\downarrow};0,m_{-\kk,\uparrow}
;0,m_{-\kk,\downarrow}\rangle\nonumber,\\
 |1,-\frac12,-\kk\rangle&\equiv&|0,m_{\kk,\uparrow};0,m_{\kk,\downarrow};1,m_{-\kk,\uparrow}
;0,m_{-\kk,\downarrow}\rangle.
\end{eqnarray}
The eigenvalues are
\begin{equation}
 \varepsilon_{1}=E_{HO}^\kk-\hbar\lambda_\kk.
\end{equation}
Note that $\lambda_\kk$ is the coefficient of the kinetic energy term of the electron
--- see (\ref {eq:Hfermi}).

\item With two electrons and total spin $\pm 1$ the total momentum of the electrons has to
vanish because of Pauli's exclusion principle. The eigenvectors are
\begin{eqnarray}
 |2,+1,0\rangle&\equiv&|1,m_{\kk,\uparrow};0,m_{\kk,\downarrow};1,m_{-\kk,
\uparrow};0,m_{-\kk,\downarrow}\rangle;\cr
 |2,-1,0\rangle&\equiv& |0,m_{\kk,\uparrow};1,m_{\kk,\downarrow};0,m_{-\kk,\uparrow}
;1,m_{-\kk,\downarrow}\rangle.
\end{eqnarray}
The corresponding energy value is
\begin{equation}
 \varepsilon_{2}\equiv E_{HO}^\kk-2\hbar\lambda_\kk+\hbar\eta.
\end{equation}

\item With two electrons and total spin zero the interaction with the harmonic
oscillators contributes. An additional label is needed to distinguish between the
two levels of each of the Jaynes-Cummings pairs. The eigenvectors are
\begin{eqnarray}
 |2,0,\kk,-\rangle&=&\sin\theta_\kk^+|1,m_{\kk,\uparrow};0,m_{\kk,\downarrow};0,m_{
-\kk,\uparrow};1,m_{-\kk,\downarrow}\rangle\nonumber\\
 &&+\cos\theta_\kk^+|0,m_{\kk,\uparrow}+1;1,m_{\kk,\downarrow};1,m_{-\kk,
\uparrow};0,m_{-\kk,\downarrow}-1\rangle\nonumber\\
 \\
 |2,0,\kk,+\rangle&=&\cos\theta_\kk^+|1,m_{\kk,\uparrow};0,m_{\kk,\downarrow};0,m_{
-\kk,\uparrow};1,m_{-\kk,\downarrow}\rangle\nonumber\\
 &&-\sin\theta_\kk^+|0,m_{\kk,\uparrow}-1;1,m_{\kk,\downarrow};1,m_{-\kk,
\uparrow};0,m_{-\kk,\downarrow}+1\rangle\nonumber\\
 \\
 |2,0,-\kk,-\rangle&=&\sin\theta_\kk^-|0,m_{\kk,\uparrow};1,m_{\kk,\downarrow}
;1,m_{-\kk,\uparrow};0,m_{-\kk,\downarrow}\rangle\nonumber\\
 &&+\cos\theta_\kk^-|1,m_{\kk,\uparrow}+1;0,m_{\kk,\downarrow};0,m_{-\kk,\uparrow}
;1,m_{-\kk,\downarrow}-1\rangle\nonumber\\
 \\
 |2,0,-\kk,+\rangle&=&\cos\theta_\kk^-|0,m_{\kk,\uparrow};1,m_{\kk,\downarrow}
;1,m_{-\kk,\uparrow};0,m_{-\kk,\downarrow}\rangle\nonumber\\
 &&-\sin\theta_\kk^-|1,m_{\kk,\uparrow}-1;0,m_{\kk,\downarrow};0,m_{-\kk,\uparrow}
;1,m_{-\kk,\downarrow}+1\rangle.\nonumber\\
\end{eqnarray}
The angles $\theta_\kk^\pm$ are given by
\begin{equation}
 \theta_\kk^\pm=\arctan\left(\frac{\omega_\kk+\omega_{-\kk}}{2\xi_\kk}\mp\frac{1}{
2\xi_\kk}\sqrt{(\omega_\kk+\omega_{-\kk})^2+4\xi_\kk^2}\right),
\end{equation}
like in the Jaynes-Cummings model \cite {JC63}.
The energies for these states are:
\begin{equation}
 \varepsilon^\pm_{2}=E_{HO}^\kk-2\hbar\lambda_\kk\pm\frac{\hbar}{2}\sqrt{
(\omega_\kk+\omega_{-\kk})^2+4\xi_\kk^2},
\end{equation}
where the plus sign holds for the first and the third, and the minus sign for
the second and the fourth class.

\item With three electrons there are again four different classes of states:
\begin{eqnarray}
 |3,+\frac12,\kk\rangle&=&|1,m_{\kk,\uparrow};1,m_{\kk,\downarrow}
;1,m_{-\kk,\uparrow};0,m_{-\kk,\downarrow}\rangle\\
 |3,+\frac12,-\kk\rangle&=&|1,m_{\kk,\uparrow};0,m_{\kk,\downarrow};1,m_
{-\kk,\uparrow};1,m_{-\kk,\downarrow}\rangle\\
 |3,-\frac12,-\kk\rangle&=&|0,m_{\kk,\uparrow};1,m_{\kk,\downarrow};1,
m_{-\kk,\uparrow};1,m_{-\kk,\downarrow}\rangle\\
 |3,-\frac12,\kk\rangle&=&|1,m_{\kk,\uparrow};1,m_{\kk,\downarrow};0,
m_{-\kk,\uparrow};1,m_{-\kk,\downarrow}\rangle
\end{eqnarray}
all with the same energy
\begin{equation}
 \varepsilon_3=E_{HO}^\kk-3\hbar\lambda_\kk.
\end{equation}

\item With four electrons:
\begin{equation}
 |4,0,0\rangle=|1,m_{\kk,\uparrow};1,m_{\kk,\downarrow};1,m_{-\kk
,\uparrow};1,m_{-\kk,\downarrow}\rangle.
\end{equation}
The energy is
\begin{equation}
 \varepsilon_4=E_{HO}^\kk-4\hbar\lambda_\kk+2\hbar\eta_\kk.
\end{equation}

\end{enumerate}

\end{document}